\documentclass{ws-ijmpe}
\usepackage[super,compress]{cite}
\usepackage{color,ulem}
\usepackage{braket}
\begin{document}

\markboth{T.~Kouno,~C.~Ishizuka,~K.~Fujio,~T.~Inakura,~S.~Chiba}{Effects of triaxiality \& pairing interaction on fission barriers of actinide nuclei}

\catchline{}{}{}{}{}

\title{Effects of triaxiality and pairing interaction on fission barriers of actinide nuclei}

\author{Taiki Kouno}

\address{Tokyo Institute of Technology, 2-12-1 Ookayama, Meguro, Tokyo 152-8550, Japan}

\author{Chikako Ishizuka\footnote{Corresponding author}}

\address{Tokyo Institute of Technology, 2-12-1 Ookayama, Meguro, Tokyo 152-8550, Japan\\
ishizuka.c.aa@m.titech.ac.jp}https://ja.overleaf.com/project/61bac26ebaa5e200a37e7ab8

\author{Kazuki Fujio}

\address{Tokyo Institute of Technology, 2-12-1 Ookayama, Meguro, Tokyo 152-8550, Japan}

\author{Tsunenori Inakura}

\address{Tokyo Institute of Technology, 2-12-1 Ookayama, Meguro, Tokyo 152-8550, Japan}

\author{Satoshi Chiba}

\address{Tokyo Institute of Technology, 2-12-1 Ookayama, Meguro, Tokyo 152-8550, Japan}

\maketitle

\begin{history}
\received{Day Month Year}
\revised{Day Month Year}
\end{history}

\begin{abstract}
We employ the density-dependent relativistic mean-field theory to study how the triaxiality and pairing interaction affect the inner fission barriers of actinide nuclei.
It was found that the triaxiality reduced the inner barriers and  improved agreement with experimental values for many actinides. However, 
about 1-2 MeV discrepancy to the experimental values
still remained for some 
of the considered nuclei.
Such a discrepancy could be made further smaller by increasing the BCS pairing strength parameter.
%
In this work, we demonstrated that adjusting the paring strength was effective in reproducing the experimental inner fission barriers as well as  ``pairing rotational energy’’ and binding energy in a consistent manner for nuclei where the effect of the triaxiality on 
the inner barriers was significant.
\end{abstract}


\ccode{25.85.-w, 
24.10.Jv, 
21.10.Gv, 
}

\section{Introduction}
Nuclear fission is a phenomenon in which a nucleus transforms into two fission fragments while emitting occasionary a few neutrons or light-charged particles. Understanding this phenomenon is essential for both applications such as nuclear technology and fundamental science such as nucleosynthesis in the cosmos.

Bohr and Wheeler’s pioneering work on nuclear fission was done based on the liquid-drop model, and there existed only one fission barrier or saddle point, corresponding to the activation energy of the chemical reaction, in the potential energy characterized by deformation coordinates of nuclei~\cite{Bohr}. 
Recent abundant experiments revealed that the potential energy of a fissioning system had typically two fission barriers which were classified as ``inner’’ and ``outer’’  fission barrier~\cite{exp}.  Nuclear fission theory had been advanced steadly by introduction of deformed single-particle orbits and a shell correction proposed by Strutinsky~\cite{Strutinsky1,Strutinsky2}, which could
reproduce such a double-hamped structure of the 
potential energy surface.  Furthermore, microscopic theories, both non-relativistic and relativistic ones, are now actively applied to the study of nuclear fission.

Among the fission barriers, the inner one is a crucial physical quantity used to determine the neutron fission cross-section of actinide nuclei since it is higher than the outer one for most of the actinides important for energy application\cite{review}.
Traditionally, the inner fission barriers have been calculated by assuming axial symmetry in many works because of the low computational cost; typical codes used in applications, such as SkyAx \cite{SkyAx} and two-center shell model \cite{tcsm}, assume the axial symmetry. 
However, the inner barriers of actinides calculated by assuming the axial symmetry ended up as an overestimation by about $2\sim5$ MeV compared to experimental values~\cite{Skyrme, Gogny}. 
On the other hand, the role of the triaxiality for the inner barrier has been theoretically examined in a few decades. Many of such works used the macroscopic-microscopic method~\cite{mm1, mm2} and non-relativistic energy density functionals based on Skyrme force and Gogny force~\cite{nr1,nr2}.
Nowadays a Chinese group also found that the triaxiality suppresses the inner barrier height by about 2 MeV using the relativistic mean-field (RMF) + point coupling model~\cite{China1, China2}. Although the influence of the triaxiality on the inner fission barriers varies from theory to theory, we cannot ignore definitely the influence brought by the triaxial deformation. 
Our previous study~\cite{Myreport} found that increasing the pairing strength could enhance the agreement of the calculated inner fission barriers with experimental data 
on the assumption of the axial symmetry.  
We evaluated the pairing strength thus obtained by using the concept of pairing rotation, which was an excellent observable to determined the pairing correlation because it could avoid calculations for odd nuclei and maintain time-reversal symmetry \cite{pairingrotation,pairingrotation2}.
Then, it is the purpose of the present study to evaluate the inner fission barriers of actinide nuclei by taking account of the triaxiality using density-dependence relativistic mean-field theory \textcolor{black}{to see how the interplay between the triaxiality and 
pairing strength can explain the systematical trends of the inner barrier of actinide nuclei.}

Our paper is organized as follows. Section 2 introduces the density-dependence relativistic mean-field theory and pairing rotation. In subsection 3.1, we first examine the changes in the inner fission barrier obtained with the assumptions of axial symmetry and triaxiality for the case of $^{240}{\rm{Pu}}$ as an example. In subsection 3.2, we show the results of a broader region of actinide nuclei. We also discuss the effects of pairing interactions under the  triaxiality on the inner barrier. Especially we address the necessity of increasing pairing strength to improve reproducibility of the calculated barriers with experimental values. Then we summarize results of our study in section 4.

\section{Method}
\subsection{Density-dependent relativistic mean-field theory}
We used two density-dependent relativistic mean-field theories in this study: density-dependent meson-exchange model (DD-ME2) and density-dependent point-coupling interaction (DD-PC1) \cite{program,DDME2,DDPC11,DDPC12}. 

Firstly, the Lagrangian in DD-ME2 is as follows \cite{DDME2};
\begin{eqnarray}
\mathcal{L}&=&\bar{\psi}(i\gamma\cdot\partial-m)\psi+\frac{1}{2}(\partial\sigma)^{2}-\frac{1}{2}m_{\sigma}^{2}\sigma^{2}-\frac{1}{4}\Omega_{\mu\nu}\Omega^{\mu\nu}+\frac{1}{2}m_{\omega}^{2}\omega_{\mu}\omega^{\mu}-\frac{1}{4}\vec{R}_{\mu\nu}\cdot\vec{R}^{\mu\nu} \nonumber\\
&+&\frac{1}{2}m_{\rho}^{2}\vec{\rho}_{\mu}\cdot\vec{\rho}^{\mu}
-\frac{1}{4}F_{\mu\nu}F^{\mu\nu}-g_{\sigma}(\rho)\sigma\bar{\psi}\psi-g_{\omega}(\rho)\bar{\psi}\gamma^{\mu}\psi\omega_{\mu}-g_{\rho}(\rho)\bar{\psi}\vec{\tau}\gamma^{\mu}\psi\cdot\vec{\rho} \nonumber\\
&-&e\bar{\psi}\gamma^{\mu}\psi A_{\mu} \label{DDME2Lagrangian}\\
\Omega_{\mu\nu}&=&\partial_{\mu}\omega_{\nu}-\partial_{\nu}\omega_{\mu},~\vec{R}_{\mu\nu}=\partial_{\mu}\vec{\rho}_{\nu}-\partial_{\mu}\vec{\rho}_{\mu},~F_{\mu\nu}=\partial_{\mu}A_{\nu}-\partial_{\nu}A_{\mu} \label{DDME2Tensor}\\
g_{i}(\rho)&=&g_{i}(\rho_{sat})f_{i}(x),~\mbox{for}~i=\sigma,\omega,~
f_{i}(x)=a_{i}\frac{1+b_{i}(x+d_{i})^{2}}{1+c_{i}(x+d_{i})^{2}} \label{DDMECouplingConstant1} \\
g_{\rho}(\rho)&=&g_{\rho}(\rho_{sat})\exp\left[-a_{\rho}(x-1)\right],~x=\rho/\rho_{sat},~\rho_{sat}=0.152{\rm{fm}^{-3}} \label{DDMECouplingConstant2}  \ .
\end{eqnarray}
Here,  $\psi$ denotes the four-component nucleon Dirac field, $\sigma,~\omega^{\mu},~{\bm{\rho}}^{\mu}$ and $A^{\mu}$ indicate scalar-isoscalar, vector-isoscalar, vector-isovector and the photon field, respectively.  The symbol $m$ designates the nucleon mass, while $m_{\sigma},~m_{\omega},~m_{\rho}$ denote the meson masses, and $g_{\sigma}(\rho),~g_{\omega}(\rho),~g_{\rho}(\rho)$ represent the density-dependent meson-nucleon coupling functions. 
The symbol $\rho_{sat}$ is the saturation density of nuclear matter.
In the relativistic mean-field approach, we ignore the fluctuation of the meson field and the negative energy components (so-called no-sea approximation). Furthermore, one of the reasons why the coupling function has density-dependence is to reproduce the nuclear matter properties calculated by Relativistic Brueckner-Hartree-Fock \cite{Brockman}.

\begin{table}[pt]
\tbl{Parameter values used in DD-ME2 \cite{program,DDME2}}
{\begin{tabular}{@{}ccccccc@{}} \toprule
&$m_{i}({\rm{MeV}})$&$g_{i}(\rho_{sat})$&$a_{i}$&$b_{i}$&$c_{i}$&$d_{i}$ \\ \colrule
$i=\sigma$&$550.1238$&$10.5396$&$1.3881$&$1.0943$&$1.7057$&$0.4421$ \\
$i=\omega$&$783.0000$&$13.0189$&$1.3892$&$0.9240$&$1.4620$&$0.4775$ \\ 
$i=\rho$&$763.0000$&$\hphantom{0}3.6836$&$0.5647$&-&-&- \\ \botrule
\end{tabular}}
\label{DDME2Table}
\end{table}

Table~1 
shows parameters in DD-ME2, which were determined to reproduce the binding energies, charge radii, and neutron radii of spherical nuclei \cite{program,DDME2}. This parameter set yields the followng nuclear-matter properties; the binding energy per nucleon is $16.24~{\rm{MeV}}$, the incompressibility is $250~{\rm{MeV}}$, and the symmetry energy at saturation density is $32.3~{\rm{MeV}}$.

Nextly, the Lagrangian in DD-PC1 is given as follows \cite{DDPC11,DDPC12};
\begin{eqnarray}
\mathcal{L}&=&\bar{\psi}(i\gamma\cdot\partial-m)\psi-\frac{1}{2}\alpha_{S}(\rho)(\bar{\psi}\psi)(\bar{\psi}\psi)-\frac{1}{2}\alpha_{V}(\rho)(\bar{\psi}\gamma^{\mu}\psi)(\bar{\psi}\gamma_{\mu}\psi) \nonumber\\
&-&\frac{1}{2}\alpha_{TV}(\rho)(\bar{\psi}\vec{\tau}\gamma^{\mu}\psi)(\bar{\psi}\vec{\tau}\gamma_{\mu}\psi)
-\frac{1}{2}\delta_{S}(\partial_{\nu}\bar{\psi}\psi)(\partial^{\nu}\bar{\psi}\psi)-e\bar{\psi}\gamma\cdot A\frac{1-\tau_{3}}{2}\psi \label{DDPC1Lagrangian} \ ,\\
\alpha_{i}(\rho)&=&a_{i}+(b_{i}+c_{i}x)e^{-d_{i}x},~(i=S,V,TV),~x=\rho/\rho_{sat},~\rho_{sat}=0.152\:{\rm{fm}^{-3}} \ .
\end{eqnarray}
Here, $m$ is the nucleon mass, while $\alpha_{S}(\rho),~\alpha_{V}(\rho),~\alpha_{TV}(\rho)$ represent the density-dependent point coupling functions. The symbol $\psi$ represents the nucleon four-component Dirac field. The DD-PC1 treats the nucleon Dirac field in the same manner as DD-ME2.

\begin{table}[pt]
\tbl{Parameter values used in DD-PC1 \cite{program,DDPC11,DDPC12}. }
{\begin{tabular}{@{}ccccc@{}} \toprule
&$a_{i}({\rm{fm}}^{2})$&$b_{i}({\rm{fm}}^{2})$&$c_{i}({\rm{fm}}^{2})$&$d_{i}$\\ \colrule
$i=S$&$-10.0462\hphantom{0}$&$-9.1504$&$-6.4273$&$1.3724$ \\
$i=V$&$\hphantom{000}5.95195$&$\hphantom{00}8.8637$&$\hphantom{00}0.0\hphantom{000}$&$0.6584$ \\
\hphantom{0}$i=TV$&$\hphantom{000}0.0\hphantom{0000}$&$\hphantom{00}1.8360$&$\hphantom{00}0.0\hphantom{000}$&$0.6403$ \\ \botrule
\end{tabular}}
\label{DDPC1Table}
\end{table}

The parameters of DD-PC1 shown in Table 
2, were determined to reproduce properties of infinite and semi-infinite nuclear matter, and the binding energies of axially symmetric deformed nuclei in the mass regions $A\approx150-180$ and $A\approx230-250$ \cite{program,DDPC11,DDPC12}. Here, the following nuclear-matter properties were obtained; the binding energy per nucleon is 16.14 MeV, the incompressibility is 250.89 MeV, and the symmetry energy at saturation density is 32.3 MeV.

In these models, the total energy is defined as $E=E_{RMF}+E_{pair}+E_{cm}$. The $E_{RMF}$ can be calculated by volume-integration of the hamiltonian density obtained from the Lagrangian density \cite{Walecka}. 
The center-of-mass energy $E_{cm}$ is described as $E_{cm}=-{{\bm{P}}_{cm}^{2}}/({2Am})$, where ${\bm{P}}_{cm}$ is the center-of-mass momentum. 
The pairing energy $E_{pair}$ is obtained as $E_{pair}=-(G/4) \mathrm{Tr}\left( \kappa^\dag\kappa\right)$, where $G$ denotes
the pairing strength and $\kappa$ the pair tensor deternined by the relativistic Hartree-Bogoliubov calculation\cite{program,pairing}. The pairing strength was determined as $G=728~ {\rm{MeV\cdot fm^{3}}}$ to reproduce 
the pairing gap in symmetric nuclear matter\cite{program}.
While there are several types of the pairing correlation, we use the simplest monopole pairing at the present calculations because complicated form of the pairing correlation introduces additional parameters and induces the further model dependence.
We used the deformed harmonic-oscillator basis in the density-dependent models. As for the details of the numerical method, please refer to a reference\cite{program}.

We imposed constraints on the axial and triaxial mass quadrupole moments to investigate the total energy dependence on the deformation covering the inner fission barrier. 
We defined the Hamiltonian as an unrestricted variation of the following function:
\begin{eqnarray}
\braket{H}+\sum_{\mu=0,2}C_{2\mu}\left(\braket{\hat{Q}_{2\mu}}-q_{2\mu}\right)^{2} \label{q1} \ , 
\end{eqnarray}
where $\braket{H}$ is total energy, while $\braket{\hat{Q}_{2\mu}}$ denotes the expectation values of the mass quadrupole operators defined as follows:
\begin{eqnarray}
\hat{Q}_{20}=2z^{2}-x^{2}-y^{2},~\hat{Q}_{22}=x^{2}-y^{2}. \label{q3}
\end{eqnarray}
 Furthermore, $q_{2\mu}$ define the constrained values of multipole moment, and $C_{2\mu}$ are the corresponding stiffness constants \cite{program,Ring}.

\subsection{Pairing rotation}
\label{Subsec:Pairing rotation}
In this work, we will improve the reproducibility of the inner fission barriers by increasing the strength of the pairing strength under the triaxial deformation. 
Generally, the pairing correlation had been evaluated in the past by using, as mentioned above, the pairing gap obtained from odd-even staggering of binding energy or neutron separation energy.  Calculation 
of the pairing gap, however, suffers from calculations for odd nuclei since it involes procedure such as blocking method, which has an ambiguity originating from the mean-field part of the theory. 

On the other hand, the pairing rotational energy can maintain the time-reversal symmetry and avoid the calculation of odd nuclei because it can be determined only by the binding energies of even-even nuclei. Therefore, in recent years, pairing rotation has attracted attention as an effective observable to determine pairing interaction.
Pair correlation breaks the U(1) gauge symmetry, and it derives the pairing rotational energy defined in the 3rd term of r.h.s. of the following equation; \cite{pairingrotation,pairingrotation2}:
\begin{eqnarray}
E(N,Z_{0})=E(N_{0},Z_{0})+\lambda_{n}(N_{0},Z_{0})\Delta N+\frac{(\Delta N)^{2}}{2\mathcal{I}_{nn}(N_{0},Z_{0})}
\label{pairingrotation} \ ,
\end{eqnarray}
where $E(N_0,Z_0)$ denotes ground-state energy for neutron number of $N_0$ and proton number of $Z_0$ nucleus, $\Delta N=N-N_{0},\ \lambda_n(N_{0},Z_0)=dE/dN|_{N=N_{0}, Z=Z_0}$ is the chemical potential, and the second-order term is the pairing-rotational energy with the pairing moment-of-inertia $\mathcal{I}_{nn}(N_0,Z_0)^{-1}=d^{2}E/dN^{2}|_{N=N_{0}, Z=Z_0}$. 
The experimental pairing rotational energy was obtained using the binding energies of even-even nuclei taken from AME2016 \cite{AME}. 

\subsection{Non-relativistic mean-field model}
\label{Subsec:non-rela}

We also performed the same calculations with non-relativistic mean-field theory for confirmaing validity and generality of results. 
\textcolor{black}{
Skyrme Hartree-Fock plus BCS pairing (SHFBCS) with SkM$^\ast$ interaction \cite{SkMs} was used. 
The pairing functional is monopole-type, smoothed constant $G$ model.~\cite{Tajima1996}
The pairing strength is obtained self-consistently by satisfying the nucleon number conservation and solving the gap equation.
The computational code we use was the one developed in Ref.~\cite{SHFBCScode}.
}

\section{Results and Discussion}
\subsection{Total Energy}
In this paper, we focus on the fifteen actinides with the charge number $Z=92-96$,
$^{232}{\rm{U}},~^{234}{\rm{U}},~^{236}{\rm{U}},~^{238}{\rm{U}},~^{240}{\rm{U}},~^{232}{\rm{Pu}},~^{234}{\rm{Pu}},
~^{238}{\rm{Pu}},~^{240}{\rm{Pu}},~^{242}{\rm{Pu}},~^{244}{\rm{Pu}},~^{242}{\rm{Cm}},\\~^{244}{\rm{Cm}},~^{246}{\rm{Cm}},~^{248}{\rm{Cm}}$,
as representative compound systems of neutron-induced fission reactions to investigate the triaxiality effects on the height of inner fission barriers.

\begin{figure}[htbp]
\centering  
\includegraphics[width=0.99\textwidth]{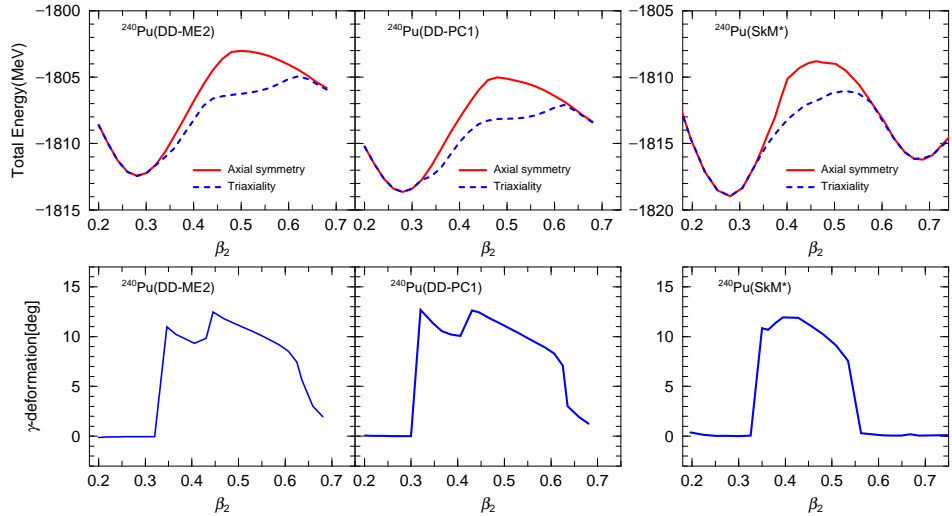}
\caption{\textcolor{black}{(Upper panels) } The $\beta_{2}$ dependence of total energy of $^{240}{\rm{Pu}}$. 
The solid and dashed lines represent axial symmetry and triaxiality, respectively. \textcolor{black}{The leftmost panel is the result with DD-ME2, the middle one is that with DD-PC1, and the rightmost one by the SHFBCS model, respectively}.\\
\textcolor{black}{~~~~~~~~~(Lower panels) The $\gamma$-deformation dependence of $^{240}{\rm{Pu}}$ on $\beta_{2}$. The results from left to right panels were calculated by DD-ME2, DD-PC1 interactions, and SHFBCS model, respectively.}
}
\label{Figure1}
\end{figure}

First, we show in \textcolor{black}{upper panels of} Fig.~\ref{Figure1} the $\beta_{2}$ dependence of the total energy of $^{240}{\rm{Pu}}$ for 
axial symmetry (solid line) and triaxiality (dashed line) cases.   Here, $\beta_{2}$ stands for the quadrupole moment.
We found that the global behaviour of the total energy seems to be common in two RMF \textcolor{black}{calculations and SHFBCS calculation}; the ground state appears at $\beta_2\simeq0.3$ in both assumptions on axial symmetry. 
On the other hand, the position of the saddle point depends on the symmetry assumption, namely, the saddle is at $\beta_{2}\simeq0.5$ in the axial symmetry cases, while the saddle is at $\beta_2\simeq0.6$ in the triaxial cases.
%
\textcolor{black}{Note that we do not performed RMF calculations for $\beta_0 > 0.7$, though there would exist a second minum like the Skyrme case. In the RMF case, we employed the harmonic oscillator basis expansion method which has less reliability for large deformation.}

We defined the height of the inner fission barrier as the difference between the total energies of the saddle point and the ground state. The inner fission barrier heights obtained from \textcolor{black}{upper panels of} Fig.~\ref{Figure1} are given in Table
3.

\begin{table}[pt]
\tbl{The inner fission barrier height (MeV) of $^{240}{\rm{Pu}}$. }
{\begin{tabular}{@{}cccc@{}} \toprule
Symmetry / model &DD-ME2&DD-PC1 & SkM$^\ast$\\ \colrule
Axial symmetry&$9.40$&$8.61$& \textcolor{black}{10.16}\\ 
Triaxiality&$7.50$&$6.53$&\textcolor{black}{7.93} \\
\botrule
\end{tabular}} 
\label{Pu240Bf}
\end{table}

From Table 
3, we found that the triaxiality makes the inner barrier height lower than that for the axial symmetry by about 2 MeV, as pointed out by the original work of DD-PC1~\cite{DDPC12}.  
%
However, the experimental height of the inner barrier of $^{240}{\rm{Pu}}$ is $5.60~{\rm{MeV}}$~\cite{exp} or $6.05~{\rm{MeV}}$~\cite{RIPL3} and still lower than those theoretical values with triaxiality listed in Table~3. 
This discrepancy suggests that theoretial fission barries must be lowered down further.
\textcolor{black}{The SHFBCS calculation has similar trend. The axail-symmetric calculation overestimates the fission barrier height by $\sim$ 4 MeV. 
Taking triaxiality improve this overestimation by 2 MeV, but still an overestimate remains.
}
We can observe a similar discrepancy between experimental inner barrier heights and theoretical heights assuming triaxiality
in other actinides, as will be shown in Fig.~\ref{Figure5} later.   It is our experience on the pairing strength as described in
the last publication\cite{Myreport} which helps to reduce the discrepancy.
%
%
In the following subsections, we will show the critical effect of the  pairing interaction to lower the inner fission barrier. Before doing so, we wish to discuss detailed feature of results for $^{240}{\rm{Pu}}$ obtained with DD-ME2 and DD-PC1 under both assumptions of  axial symmetry and triaxiality.

\begin{figure}[htbp]
\centering  
\includegraphics[scale=0.5]{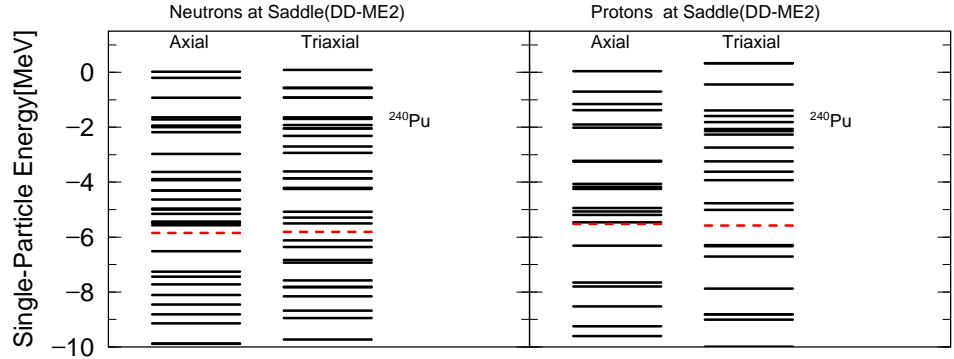}
\caption{Neutron (left) and Proton (right) single-particle energies of $^{240}{\rm{Pu}}$ at saddle point assuming axial symmetry and triaxiality calculated with DD-ME2. The horizontal dashed linse denote the Fermi energy (or chemical energy in BCS theory). }
\label{Figure2}
\end{figure}

\begin{figure}[htbp]
\centering  
\includegraphics[scale=0.5]{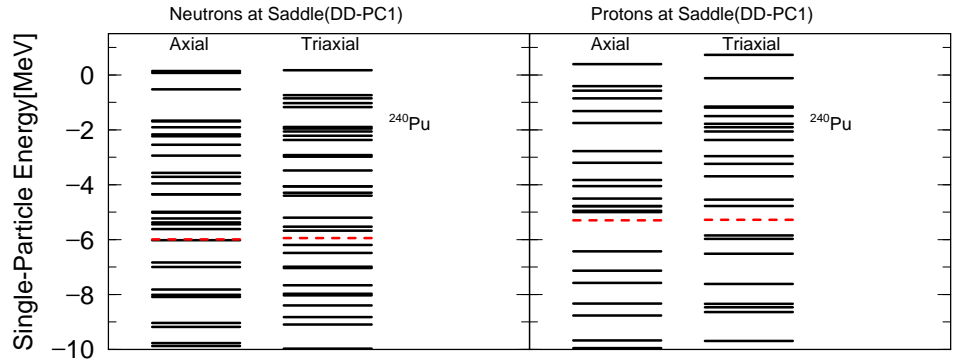}
\caption{Neutron (left) and Proton (right) single-particle energies of $^{240}{\rm{Pu}}$ at saddle point for axial symmetry and triaxiality calculated with DD-PC1. The horizontal dashed linse denote the Fermi energy (or chemical energy in BCS theory). }
\label{Figure3}
\end{figure}

One of the reasons which brought the difference of the inner fission barrier heights obtained by the axial symmetry and the triaxial asymmetry is the level density at the saddle. Figures~\ref{Figure2} (with DD-ME2) and~\ref{Figure3} (with DD-PC1) depict proton and neutron single-particle \textcolor{black}{levels} of $^{240}{\rm{Pu}}$ at the saddle point under the assumption of axial symmetric deformation and triaxial deformation. 
\textcolor{black}{The neutron Fermi energies in $^{240}$Pu with DD-ME2 are almost  the same in two cases; $E_{F,n}=-5.85$ MeV for axial symmetric case and $E_{F,n}=-5.81$ MeV for triaxial case. There are seven levels in the energy range of $E_{F,n} \pm 1$ MeV for axial symmetric, while five levels lie for triaxial cases. Similar trends are seen in proton levels and also in those with DD-PC1 (Fig.~\ref{Figure3}).
For protons, the levels around the Fermi energy are sparser for the triaxial case than in the axial case.}
We can see that the triaxial deformation yields a smaller level density than the axial symmetry deformation at the Fermi energy.

\textcolor{black}{Lower panels of Fig.~\ref{Figure1}} represents the $\gamma$-deformation of $^{240}{\rm{Pu}}$ as a function of $\beta_{2}$. 
%
 %
The shape near the ground state is axially symmetric, while that near the saddle point ($0.35<\beta_{2}<0.6$) is triaxial. 
%
%
 It is known that the $\gamma$-deformation near the saddle point is almost universal in actinides~\cite{Tri1} at $\gamma \simeq10^{\circ}$.
\textcolor{black}{
Another noticeable feature of the relation between the $\gamma$-deformation and $\beta_2$ is the sudden jump around $\beta_2=0.3$. When $\beta_2$ is fixed, there are two local minuma at $\gamma=0$ and $\gamma\simeq 10^{\circ}$.
These two states have close energies, and the lower-energy state chages at $\beta_2 \sim 0.3$. 
This transition causes the discontinuity of $\gamma$ deformation. 
Similar discontinuities in multidimensional potential energy surfaces of actinides nuclei are reported~\cite{Zdeb,Lau,Flynn}.
}  


\subsection{Effect of triaxiality and pairing interaction for inner fission barrier}
The previous subsection picked out $^{240}{\rm{Pu}}$ as a representative case and showed various features with axial-symmetry/triaxiality. In the former part of this subsection, we display the inner fission barriers of actinides, including $^{240}{\rm{Pu}}$, as a function of the mass of a fissioning nuclei. It suggests that we have to 
reduce the calculated fission barriers, even if we consider the triaxiality, to reproduce the experimental data not only in $^{240}{\rm{Pu}}$ but also in other actinides. Then, in the latter part of this subsection, we focus on the influence of the pairing interaction on the fission barrier heights. Our previous study for the axial-symmetric cases showed that it could lower the heights of inner fission barriers .

\begin{figure}[htbp]
\centering  
\includegraphics[scale=0.35]{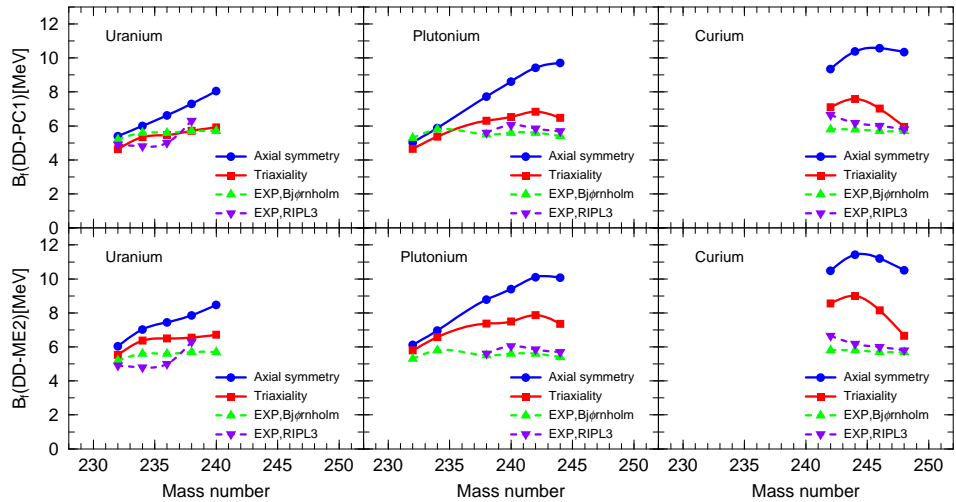}
\caption{Comparison of the calculation results of inner fission barrier by axial symmetry and triaxiality with experiment values for Pu isitopes\cite{exp,RIPL3}. }
\label{Figure5}
\end{figure}

%
Figure~\ref{Figure5}
compares the experimental inner-fission-barrier heights~\cite{exp,RIPL3} of actinides with those using two interactions (DD-ME2 and DD-PC1) in axial-symmetric and triaxial cases. 
These results show that the inner fission barrier is generally about $0.5\sim5$ MeV lower than the axial-symmetric case when we impose the triaxiality.
We also found that the reduction width depends on the number of neutrons and protons. 
However, the barrier heights of a few nuclei (e.g. $^{240}{\rm{Pu}}$) are still overestimated by about 2 MeV compared to the experimental values, even assuming the triaxiality.

\begin{figure}[htb]
\centering  %
\includegraphics[scale=0.4]{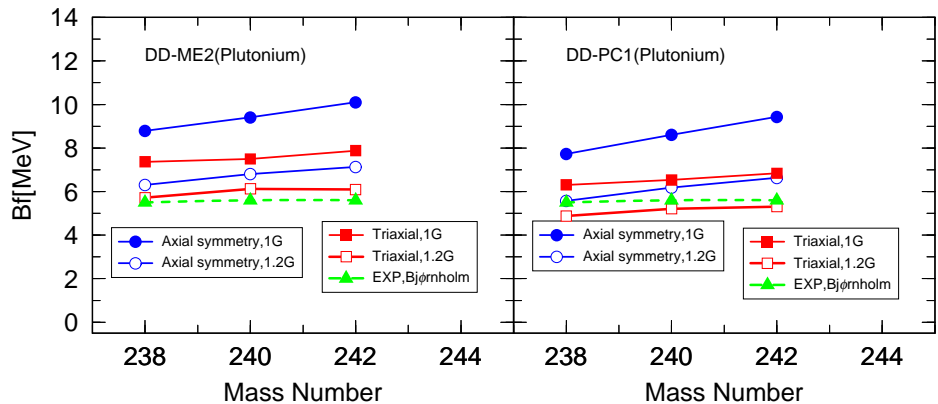}
\caption{The changes of inner fission barrier by each pair strength on axial symmetry and triaxiality. The result of the left panel was calculated by DD-ME2, the right one was calculated by DD-PC1.}
\label{Figure6}
\end{figure}

%
The above problem of overestimating experimental values can be solved by increasing the pairing strength. Here, we define the pairing strength as a single multiplier parameter $\alpha$ such as $G\rightarrow\alpha G$.
We show the results of inner fission barriers with unchanged pairing strength ($\alpha=1.00$) and with changed pairing strength by $+20\%$ ($\alpha=1.20$) in Fig.~\ref{Figure6}. Here, experimental values $B_{f}(\mbox{Exp})$ were taken from \cite{exp,RIPL3}. In addition, as an example, we selected three Pu-isotopes, where the fission barriers with triaxiality still deviate from the experimental values by more than $1$~MeV.
Such a combination of triaxiality and pairing interaction, both models (DD-ME2 and DD-PC1) can drastically improve the reproducibility of the experimental fission barriers in Pu-isotopes by increasing the pairing strength regardless of axial symmetry (blue unfilled circles) or triaxiality (red unfilled squares) as shown in Fig.~\ref{Figure6}.

Thus, one of the reasons why the height of the inner fission barrier changes depending on the strength of the pairing force is because the single-particle level densities around the Fermi surface at the saddle point is higher than the ground state.
It leads to the easier occurrence of interactions between particles at the saddle than at the ground state. Thus, the particles at the saddle can gain larger paring energy to the negative direction. 
Then stronger pairing interaction can reduce the fission barrier height.
For a more detailed discussion and analysis, see the reference~\cite{RMFBCS}. In the referred article, the authors mentioned the case only with axial symmetry; however, there is no significant difference in a physical picture between axial symmetry and triaxiality.

The primary effect of the pairing interaction on the fission barrier, i.e. reduction of the barrier height, is common in both symmetries. However, the strength of that effect depends on the symmetry, axial-symmetry or triaxiality, as shown in Fig.~\ref{Figure6}. The $B_{f}$ change due to the pairing strength $\alpha$ is larger in the axial-symmetric case than in the triaxial case because the level density assuming axial-symmetry is denser around the Fermi energy than in the triaxial case. 
It suggests that the sensitivity of the $B_{f}$ on $\alpha$ comes from the difference in the level density structures between these two symmetries.

The above discussion on Pu-isotopes concludes that the larger pairing strength can lower the inner fission barrier. 
We can expect that such a combination of the triaxiality and the pairing interaction would effectively reduce the fission barrier height in the other actinides. However, the best parameter search on $\alpha$ requires a very high computational cost, which will be our following paper's subject.

\begin{figure}[htb]
\centering  %
\includegraphics[scale=0.28]{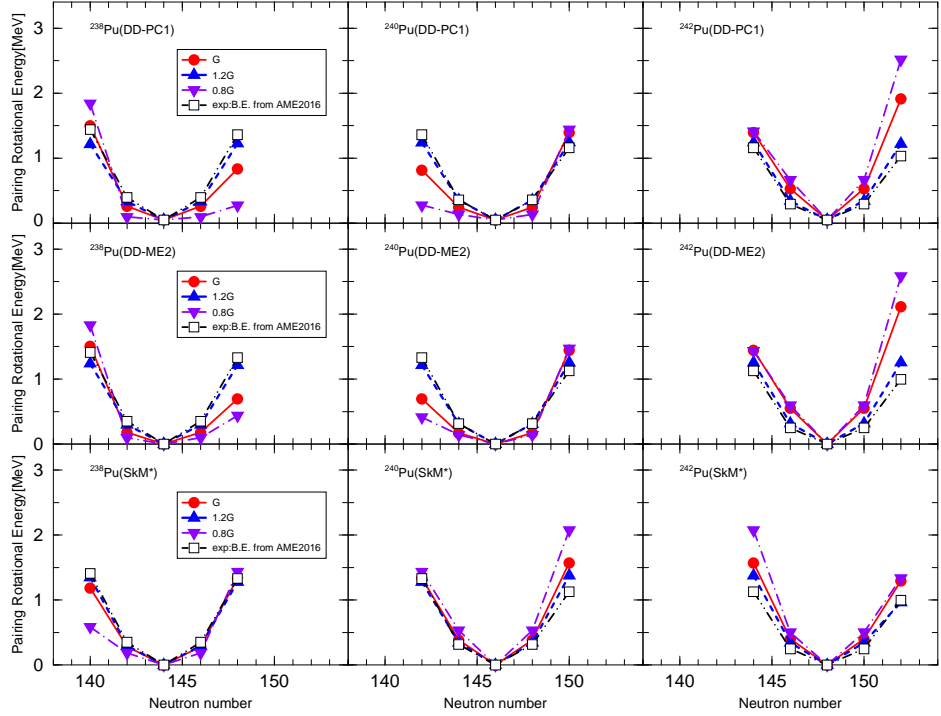}
\caption{The changes of pairing rotational energy by each pair strength. 
\textcolor{black}{The result was calculated by DD-ME2 (upper), DD-PC1 (middle), and SkM$^\ast$ (lower).}}
\label{Figure7}
\end{figure}

%
Finally, we refer to why we set the pairing strength of  Pu-isotope as $\alpha$=1.2. We adjusted $\alpha$ to reproduce the experimental value of the pairing rotational energy. Eq.(9) states that the pairing rotational energy depends only on the ground state energy, which is generally the same in axial-symmetric and triaxial cases, as shown in Fig.~\ref{Figure1}, 
\textcolor{black}{i.e., $\gamma=0$ at the ground state.}
It means that the pairing rotational energy with the triaxiality is the same as that with axial symmetry. 
Figure~\ref{Figure7} shows the resulting pairing rotational energy with DD-PC1 \textcolor{black}{(uppermost panels)} and DD-ME2 \textcolor{black}{(middle panels)} in the Pu isotopes. These results reveal that making pairing strength stronger by $+20\%$ (blue filled triangles) is the most reasonable among three cases ($0.8G, G, 1.2G$) to explain the experimental value (white unfilled squares) of pairing rotational energy.
\textcolor{black}{The SHFBCS calulations (lowermost panels) also show that the pairing strength with $1.2G$ repoduces the pairing rotational energies better than the others, while its differeces are small.  Therefore, we can notice herre that the effect of the pairing on the nuclear properties has a strong model dependence.}
%
Similar results are obtained in Uranium isotopes.
Therefore, 
\textcolor{black}{we can draw an important conclusion that the pairing strengths
in the actinide region are not subject to those determined by analysis for lighter nuclei, but rather, they should be determined systematically by taking account of the nuclear binding energy, fission barrier and pairing rotational energy simultaneously.   For this aim, we need to consider both of the triaxial deformation and pairing interaction in a consistent manner.}

\section{Summary}
We systematically investigated the height of the inner fission barrier of actinide nuclei based on the density-dependence relativistic mean-field theory (DD-ME2 and DD-PC1), \textcolor{black}{with taking account of the triaxial deformation}.
We found that in both the DD-ME2 and DD-PC1 parameter sets, \textcolor{black}{triaxial deformation} 
could reduce the inner fission barrier \textcolor{black}{height} compared to \textcolor{black}{ones in axial symmetric deformed cases} by about $0.5\sim5$ MeV.
However, we also found that the fission barrier height with triaxiality still deviates from the experimental value in some nuclides, for instance, $A\geq238$. For such nuclides, we increased the pairing strength by $20\%$ to improve reproducibility of the barrier \textcolor{black}{height.} 
That strength also provides more realistic pairing rotational energy, 
\textcolor{black}{which is an observable sensitive to the pairing correlation.} Thus, we obtained a consistent picture of the fission barrier and the pairing rotational energy. 
%
%
\textcolor{black}{We obtained similar results with use of non-relativistic theory, Skyrme-HF plus BCS pairing.
The important conclusion we could draw from the present work is that the pairing strengths
in the actinide region are not subject to those determined for lighter nuclei, but rather, they should be determined systematically by taking account of the nuclear binding energy, fission barrier and pairing rotational energy simultaneously.   For this aim, we need to consider both of the triaxial deformation and pairing interaction in a consistent manner.}

\section*{Acknowledgements}

This study is supported by Grands-in-Aid for Scientific Research (B) No. 21H 01856 from Japan Society for the Promotion of Science.


\end{document}